# How to Measure Cyber Resilience of an Autonomous Agent: Approaches and Challenges

Alexandre K. Ligo[3,2], Alexander Kott[1], Igor Linkov[2]

[1] U.S. Army Research Laboratory, Adelphi, MD 20783, USA
[2] Engineer Research and Development Center, US Army Corps of Engineers, Concord, MA 01742, USA
[3] University of Virginia, Charlottesville, VA 22904, USA
alexander.kott1@us.army.mil

**Abstract.** Several approaches have been used to assess the performance of cyber-physical systems and their exposure to various types of risks. Such assessments have become increasingly important as autonomous attackers ramp up the frequency, duration and intensity of threats while autonomous agents have the potential to respond to cyber-attacks with unprecedented speed and scale. However, most assessment approaches have limitations with respect to measuring cyber resilience, or the ability of systems to absorb, recover from, and adapt to cyber-attacks. In this paper, we provide an overview of several common approaches, discuss practical challenges and propose research directions for the development of effective cyber resilience measures.

## 1 Introduction

Cyber resilience has emerged as a major topic in cyber-physical systems, but measuring it objectively remains challenging. Of particular interest is the use of autonomous agents in defending systems from cyber-attacks, and the degree at which those agents are able to absorb, recover from and adapt from disruptions. However, the cyber resilience of a system can be appreciated only when adequate resilience measures are defined and implemented, which are currently missing in the literature.

This paper is a discussion of the main approaches and challenges in assessing cyber resilience of autonomous agents. We describe the main features of common approaches in cyber-physical systems and discuss limitations and challenges in measuring cyber resilience under each approach.

Cyber resilience is the ability of a system to recover or regenerate its functionality after a cyber-threat materializes and produces a degradation to the system's performance [1]. Building cyber-resilient systems is important in warfare, but the impact of recent attacks on the availability, integrity and confidentiality of systems in healthcare, utilities, logistics, etc. show the importance of cyber-resilience in civil systems as well. Assuming an attack with functionality degradation, a resilient system will restore functionality with speed commensurate with the mission or goals and will adapt in such a way that future disruptions will have relatively less impact on degrading and restoring functionality. It is generally naïve to assume that a system's robustness is such that functionality impairment following an attack is zero. However, a resilient system would degrade slowly and recover its previous functionality relatively fast. Therefore, level and speed of degradation, recovery, as well as adaptation must be measured somehow.

The fast-paced development of autonomous agents (either attackers or defenders) implies that designing cyber-resilient systems and assessing their performance becomes increasingly important [2]. For example, vehicles in the streets and in the battlefield have become progressively more connected and automated, which may make them accessible and profitable targets for swarms of automated attackers. Human defenders will not be able to respond with the necessary speed and scale, and therefore automated defenses will play a crucial role to absorb, recover from, and adapt to disruptions.

The concept of cyber resilience is not different from the general definition of resilience in other disciplines, which integrates ideas and definitions from disparate fields such as social, psychological, organizational, engineering and ecological views [3]. One general definition of resilience is that offered by the National Academy of Sciences which includes four abilities: plan or prepare for, absorb, recover from, and adapt to known threats [4], [5].

However, there are specificities in cyber resilience that make it hard to apply established knowledge from other areas. One challenge is to define and measure the critical functionality of a system and its impact on missions or processes [6]. Resilience assessment refers to the evaluation of a system's ability to absorb, recover from and adapt to disruptions. Ideally, proper assessment of cyber resilience includes functionality measurement, i.e. its quantification in an objective way. This can be difficult to realize in practice. For example, the components of a cyber-physical system may present dependencies that are relatively complex or vary over time. Or the system functions are multidimensional and include availability, confidentiality and integrity [7], meaning that it is important not only to keep the system operating but also to prevent unauthorized access. These difficulties frequently result in assessments based on subjective estimates rather than objective measures.

The rest of the paper is organized as follows. In Section 2, we describe common approaches to the assessment of systems that may involve the measurement or estimation of cyber resilience, and we discuss current challenges of each approach. Section 3 discusses a framework to measure cyber resilience and ways in which it can be improved. In Section 4 we conclude the paper and propose directions for future research.

## 2   Approaches and Challenges

This section starts with qualitative assessments of cyber resilience, followed by quantitative but subjective approaches based on expert estimates. We then discuss characteristics and challenges of more resource-intensive approaches such as modeling and simulation, wargaming, and red teaming.

### 2.1   Qualitative assessments

Some approaches to assess cyber resilience are based on qualitative evaluations of system features in several dimensions that may include resilience. One example is the cyber resilience matrix [4], which is a holistic framework of resilience metrics that cover all four resilience abilities: *plan/prepare* to keep services available and assets functioning during a disruptive event, *absorb* the disruption and maintain the most critical asset function and service availability, *recover* all asset function and service availability to their pre-event functionality, and *adapt* protocols and configuration of the system, personnel training, or other aspects to become more resilient. In the cyber resilient matrix each of those abilities have associated cyber system metrics on four domains, as shown in Table 1: *physical* resources and the capabilities and the design of those resources, *information* and information development about the physical domain, *cognitive* use of the information and physical domains to make decisions, and the *social* structure and communication for making cognitive decisions [4], [8].

While qualitative assessments such as the proposed in [4] offer guidance in the development of measures for cyber resilience, quantitative measures are not provided in a way that allows for objective assessment or comparison of resilience between different systems. Rather, comparative measures are left to be defined in a case-by-case basis.

Besides, autonomous cyber defenders generally generate logs and other forms of useful quantitative data that would be discarded in purely qualitative approaches.

**Table 1.** The cyber resilience matrix – reproduced from [4] with permission

| Plan and prepare for | Absorb | Recover from | Adapt to |
|---|---|---|---|
| **Physical** | | | |
| (1) Implement controls/sensors for critical assets | (1) Signal the compromise of assets or services | (1) Investigate/repair malfunctioning controls / sensors | (1) Review asset and service configuration in response to recent event |
| (2) Implement controls/sensors for critical services | (2) Use redundant assets to continue service | (2) Assess damage | (2) Phase out obsolete assets and introduce new assets |
| (3) Assessment of network structure and interconnection to system components and to the environment | (3) Dedicate cyber resources to defend against attack | (3) Assess distance to functional recovery | |
| (4) Redundancy of critical physical infrastructure | | (4) Safely dispose of irreparable assets | |
| (5) Redundancy of data physically or logically separated from the network | | | |
| **Information** | | | |
| (1) Categorize assets and services based on sensitivity or resilience requirements<br>(2) Documentation of certifications, qualifications and pedigree of critical hardware and/or software providers<br>(3) Prepare plans for storage and containment of sensitive information<br>(4) Identify external system dependencies (i.e., Internet, electricity, water)<br>(5) Identify internal dependencies | (1) Observe sensors for critical services and assets<br>(2) Effectively and efficiently transmit relevant data to responsible stakeholders/ decision makers | (1) Log events and sensors during event<br>(2) Review and compare systems before and after event | (1) Document incident's impact and cause<br>(2) Document time between problem and discovery/discovery and recovery<br>(3) Anticipate future system states post-recovery<br>(4) Document point of entry (attack) |
| **Cognitive** | | | |
| (1) Anticipate and plan for system states and events<br>(2) Understand performance trade-offs of organizational goals<br>(3) Scenario-based cyber wargaming | (1) Use a decision making protocol determine when event is considered ''contained''<br>(2) The ability to evaluate performance impact to determine if mission can continue<br>(3) Focus effort on identified critical assets and services<br>(4) Utilize applicable plans for system state when available | (1) Review critical points of physical and information failure in order to make informed decisions<br>(2) Establish decision making protocols or aids to select recovery options | (1) Review management response and decision making processes<br>(2) Determine motive of event (attack) |
| **Social** | | | |
| (1) Identify and coordinate with external entities that may influence or be influenced by internal cyber attacks (establish point of contact)<br>(2) Educate/train employees about resilience and resilience plan<br>(3) Delegate all assets and services to particular employees<br>(4) Prepare/establish communications<br>(5) Establish a cyber-aware culture | (1) Locate and contact identified experts and resilience responsible personnel | (1) Follow resilience communications plan<br>(2) Determine liability for the organization | (1) Evaluate employee communications preparedness and effectiveness<br>(2) Assign employees to critical areas that were previously overlooked<br>(3) Stay informed about latest threats and protection/share with organization |

### 2.2 Probabilistic estimates by experts

Some approaches use probabilistic assessments from subject matter experts as a way to replace purely qualitative judgments with more objective estimates. Such approaches may be used when historical data on emerging threats or technologies are not available, or there might be the endless possibilities in terms of forms of attacks, types of vulnerabilities, and types and magnitudes of consequences that may lead some to consider data-driven methods too complex and rather adopt subjective expert estimates.

The methods we refer to as expert assessments are those where human specialists provide estimates about the probability that an event of interest (e.g. data breach) will occur within a given period of time (e.g. one year), or the magnitude of the impact caused by the event (e.g. financial loss resulting from the data breach), or a combined estimate of likelihood and impact. Estimates may be expressed either as a quantity (e.g. expected value of the loss from a threat next year, in dollars), or a level in an ordinal scale (high/medium/low, likely/neutral/unlikely, etc.).

One popular (and controversial) example of probabilistic assessment by experts is the risk matrix. It consists of a table where columns represent an ordinal scale of magnitude of impact of risk (e.g. low, medium, high, and very high) and rows represent an ordinal scale of probability of risk occurrence as illustrated in **Table 2**. Each cell of the table will refer to a particular combination of impact and probability, and known risks are placed into the matrix according to expert assessment of the probability and impact of each risk, which should help visualize and prioritization. While the risk matrix is popular for its simplicity and visual appeal, it has been criticized for a number of flaws such as poor resolution of the matrix, error and ambiguity in judgment, etc. [9].

**Table 2.** A 4x4 risk matrix

| Impact  Probability | low | medium | high | very high |
|---|---|---|---|---|
| low | cyber risk j | | cyber risk i | |
| medium | | | | cyber risk z |
| high | | cyber risk k | | |
| very high | | | | |

Other approaches aim to replace the risk matrix with tools that refine the assessment of experts in more systematic and quantitative ways. For example, [10] have proposed a decomposition approach. Rather than estimating impact and likelihood altogether and with ordinal scales as above, experts estimate a probability value (between 0 and 100%) and separately estimate a likely range for the impact (e.g. a 90% confidence interval for the loss in dollars). Another possibility for decomposition is to split risks with highly uncertain events (e.g. overall data breach) into more manageable components (e.g. breach of particular server or application). However, the approach in [10] is still based on subjective estimates by experts that are subject to several challenges. First, cognitive issues such as overconfidence and anchoring typically include significant bias that may result in wildly inaccurate estimates. (Even though bias can be eliminated by calibration training, it is unclear how long an expert is able to provide accurate estimates before recalibration is needed.) Second, the approach in [10] lacks resilience abilities such as recovery and is actually a framework rather than proposal of concrete measures, which definition is apparently left for each organization.

Nevertheless, expert assessments may be seen as an exploratory first step within a more comprehensive approach that include quantitative measures in later steps. [10] view expert estimates as the "prior" of Bayesian approaches in which data-driven measurements are used to update "posterior" probabilities and impacts of cyber-attacks. Such a Bayesian approach could be extended to measure all cyber resilience abilities.

## 2.3 Modeling and Simulation

Modeling and simulation are frequently used to assess the impact of cyber-attacks on business or military missions. Such approaches can be characterized in two general phases. The modeling phase consists in creating a representation of the business or mission processes and dynamics, behavior of users, attackers and defenders, functions of cyber infrastructure and systems, as well as dependencies between the aforementioned components. The second phase is a simulation of known attacks and their impact on the business or mission considering the model components and dependencies.

[6] describe two case studies, one in the utilities sector (power grid) and another in the military. On the power grid study, mission impact was assessed by modeling and simulating intrusion detection in a water and energy network. An important finding was that modeling of dependencies is prohibitively expensive in complex, distributed systems. Hence, the project included a tool to automate learning of network dependencies based on network traffic, and then inferring higher level information about the business.

The second case study explored modeling and simulation of cyber-attacks on a military system. Graph models of dynamic impacts across multiple layers were developed in the operation and cyber-domains, including the mission, adversary, infrastructure, cyber defender and attack impact. The attack impacts were categorized in confidentiality, integrity or availability types. Stochastic discrete event simulation was executed to assess process dynamics and decisions following cyber-attacks. The measures developed in this study are related with mission accomplishment (rather than technical performance of cyber components) and are mission-specific, e.g. number of mission plans developed and the number of flights flown. Moreover, measurements included resilience. For example, one attack degraded systems performance by 50% but did not cause significant impact on the mission metrics, indicating high ability to absorb.

Those examples illustrate that modeling and simulating complex systems may provide a safe environment to assess cyber resilience, but that approach poses several challenges, First, it is an expensive and time-consuming exercise. Second, modeling is always a limited and imperfect representation of reality. For example, attackers and defenders should be modelled with human-like bias and imperfections, or other limitations in case of autonomous agents. Modeling of cyber-attacker's behavior might be promising but it is a "rather immature" research area [6]. Third, regardless of the level of detail a simulation model embeds strong assumptions about the attributes of the components and dependencies between them, and only accounts for threats and dependencies that are known. Automated tools might help discover unknown dependencies, but their validity is unclear. (They produced promising results in the power grid case study, they were considered unreliable for the military study). Finally, the approach of modeling and simulation does not automatically resolve the need for a formal definition of the cyber-physical mission or business process, including objective and quantitative measures for cyber resilience. Even the most realistic simulation will need adequate measures to assess the abilities to absorb, recover and adapt the system's functionality.

## 2.4 Wargaming

In contrast to human agents, autonomous cyber-defenders (and their attacking counterparts) are generally expected to act rationally in order to maximize a payoff, such as maximizing absorption or recovery in case of the defender or maximizing disruption or damage in the eyes of the attacker. Hence, game-theoretical approaches such as wargaming may fit the behavior of autonomous cyber-defenders and attackers. One example is offered in [11], which provides a formulation of the wargame that includes attacker and defender benefits, the probabilities of the attacker breaching defense layers, strategies, costs and utility for attacker and defender.

A wargame is arranged to produce a simulated but realistic experience of cyber-attacks and defenses. In this sense, wargaming is similar to modeling and simulation (see 2.3). The game may include individuals from several areas of a business or functions of a military unit, who meet for a predefined period (hours to days) to replicate the impact of cyber-attacks on technical functions or business processes. Some participants play the role of cyber-defenders (the "blue team" in wargaming jargon), while others assume the role of attackers (the "red team"). The general purpose is for the blue team to evaluate and select strategies against cyber-attacks. To accomplish that, defenders have to make assumptions about attack strategies, such as which are most likely, profitable to the red team, or damaging to the defender. Game theoretical methods are commonly used by the defenders to anticipate rational attack strategies.

The similarities of wargaming with modeling and simulation approaches extends to its challenges. Wargaming is typically a one-time, expensive arrangement that requires several, if not dozens of members of the organization. Also, it implies that the defender makes strong assumptions about the attacker strategies, behavior and resources. However, real-world agents (human or automated) have imperfect information about adversaries. Besides, in complex cyber-physical system the space of attack and defense strategies may be prohibitively large. Hence, the computation of optimal defense strategies may be impractical, leading to sub-optimal solutions. Moreover, cyber wargaming, as proposed in [11], requires that experts estimate payoffs and probabilities. Obtaining expert estimates may be a difficult and expensive process that can result in subjective, inconsistent, and/or unreliable results as discussed in Section 2.2. Finally, research is needed to develop specific measures of cyber resilience abilities (absorb, recover and adapt). Only with those measures will it be possible to develop wargames that result in quantifiable improvement in the cyber resilience of autonomous defenders.

## 2.5 Red teaming and Penetration testing

Cyber wargaming is not a universally-defined term [11] and sometimes is used interchangeably as red teaming. Both have been used in military planning and decisions for a long time [12], and both are used to denote antagonistic teams (red and blue) that compete with the purpose to evaluate the most effective defense strategies against known cyber-attacks. On the other hand, some authors, e.g. [13] make a distinction between wargaming where groups simulate attack and defense strategies in several possible settings (table-top, computational, etc.) while red teaming refers to cyber-attacks and defenses on real systems (either in live or test environments). For the sake of discussion, we adopt such a distinction between wargaming and red teaming here.

Besides, red teaming has similarities with penetration testing, which also refers to a group of skilled individuals engaged in attacking a cyber system. However, the main difference is on purpose. A red teaming exercise is objective-oriented, i.e. the red team has a specific goal such as obtaining a confidential dataset from the employee's database. In contrast, penetration testing typically focus on a specific system or application and involves a relatively broader exploration of all vulnerabilities [14].

Regardless of the distinctions between those game-like approaches, red teaming and penetration testing share the challenges mentioned in 2.4 such as the high cost of the necessary equipment and personnel, and the difficulty to map degradations in system functionality into the impact of that system's mission or business goal. Moreover, game realism comes with a trade-off. Penetration tests that are carried out on live systems are generally constrained to limit the disruption of critical functionality [15]. For those reasons, red teaming and penetration tests are typically one-shot exercises (and therefore do not enable statistical inference of outcomes) and are highly dependent on the technical skills and ingenuity of the red teams. Perhaps more importantly, those approaches concentrate on penetration and detection and do not cover other cyber resilience abilities such as recover from and adapt to disruptions.

## 3 Measuring cyber-resilience as the "Area Under the Functionality Curve"

### 3.1 Cyber resilience as integral of functionality over time

As mentioned in the previous sections, several challenges are associated with common approaches that have the potential to help estimate or measure cyber resilience.

We now turn to the specific challenge of measuring cyber resilience level and dynamics, i.e. the absorb and recover abilities in particular. In other words, we are interested not only in how much the functionality of a system inevitably degrades after a cyber-attack, but also how much and how long it takes to recover functionality. This is represented in **Fig. 1** as a plot of a system's functionality over time. In this case, the area under the functionality-time curve gives a measure of cyber-resilience [1]. For example, if two systems A and B produce the same level of functionality and are subjected to the same cyber-attack, the resilience after a period $T_m$ of system A is greater than that of B if the area under the curve is greater, i.e. either A recovers to a higher functionality level than B after $T_m$, or A recovers to the same level but faster than B.

**Fig. 1** also illustrates that the integral of the functionality curve as a measure of resilience accounts for the response to multiple (or even continuous) cyber-attacks over time. If during the period of interest $T_m$ a system is subjected to more than one cyber-attack, the area accounts for the compound effect of the multiple attacks and defenses.

While this approach to measure resilience has been applied on other disciplines [16], [17], like other approaches the area under the functionality curve over time has several challenges. First, the quantification of "functionality" of cyber-physical systems is not trivial, especially when autonomous attackers and defenders are involved and imply multiple or continuous attacks and defenses [2]. Another complication is when the critical functionality of a system varies over time. For example, say a mission requires that the system executes a function A in the first half, and then another function B. If an attacker impairs only one of those functions, it is not straightforward to determine what the degradation on the mission is. Besides, the "area" measure depends not only on the system's resilience but also on the intensity of the attack. For example, say a system is attacked and recovers completely after a certain period. Later, the same system suffers another attack that is either more intense or longer than the first one. In this example, the same system will have a second measure of resilience that is smaller than the first. Hence, the area measure actually confounds resilience with effectiveness of the attack.

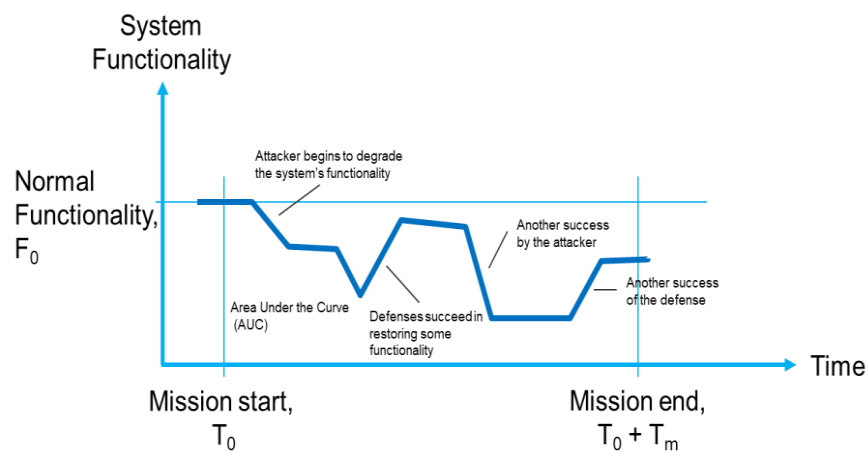

**Fig. 1.** Representation of resilience as the area under the curve of system functionality over time.

The following are possible improvements for cyber resilience measurement.

### 3.2 Mission accomplishment instead of technical functionality

As noted previously, the purpose of a cyber-physical system is better captured by measuring its accomplishment of the mission or business goal rather than indicators of technical functionality like number of users served or speed of transactions. Measuring cyber resilience through mission accomplishment would account for the importance of the system and its dependencies. For example, an attack that slows or halts a system without significantly affecting the mission or process it supports has less of an impact on resilience than attacks that effectively disrupts the mission or business.

However, if defining and measuring functionality of a cyber system is already a challenge, measuring mission accomplishment may be relatively even more difficult. Dependencies that may be indirect or interrelated in complex ways make mission or business impact challenging to identify and to quantify.

### 3.3 Adversary effort instead of time

Instead of measuring cyber resilience as the integral of functionality over time, an alternative measurement could be the integral of functionality over cumulative adversary effort (i.e. replace time with adversary effort in **Fig. 1**). Such a measure would resolve the problem that the "area" depends both on resilience and on attack effort. With effort in the horizontal axis, the measurements of resilience of different systems can be compared by using a fixed amount of effort instead of a fixed mission duration. Such a measurement should have a proper definition for the special case of zero adversary effort during a period of time. In this case a naïve formulation could result in a measure of resilience of zero, which is misleading. A possible solution is that cyber resilience for no adversary effort is *undefined*, meaning that cyber resilience is not a meaningful quantity in the absence of a threat. Nevertheless, this is subject for further discussion.

## 4 Conclusions

We discussed challenges with common approaches used in the assessment of cyber resilience. In particular, many approaches do not go beyond the evaluation of the likelihood and impact of risks, meaning they rarely address abilities particular to cyber resilience such as recover to and adapt from disruptions. Comprehensive resilience measures are especially important when autonomy is taken into consideration. Autonomous cyber-attackers and defenders will frequently and continuously interact, producing a dynamic effect on the functionality of a system that must be measured somehow.

We argue that research is needed to develop novel measurements of cyber resilience. Different approaches (probabilistic, simulation, etc.) may be combined with the idea of capturing the dynamics of resilience as the integral of functionality over time. However, autonomous agents in cyber-physical systems demand improvements in measurement. First, we need a proper (and ideally generalizable) way to define functionality. Second, improvements are needed to simplify the assessment of how the degradation of attacks on system functionality impacts the mission or business goals. Finally, we shall define measures of cyber resilience that are normalized over the effort of the adversary.

One possible direction of research may include mimicking the resilience of biological systems. Distributed autonomous agents may specialize in measuring degradation of functionality in individual components rather than trying to measure global impact. Likewise, those distributed agents would engage in responding and recover only the components they are responsible for, playing the role of cyber "antibodies", similarly as IBM envisioned in [18]. In addition, the agents might send measurements to a central "brain" which would be capable the estimate the overall change in functionality and its mission impact.

This article does not fulfil the needs of new cyber resilience measures. Instead, we are inviting the community of cyber researchers and practitioners to engage in conversation and exploration. Clearly, we cannot reliably improve what we cannot measure. All sciences and engineering fields blossomed only when measurement tools were developed and accepted as common language. No technical discipline has achieved a de-gree of maturity without developing measurement tools and methods, in order to measure objectively, rigorously, and quantitatively the attributes of phenomena occurring in the systems of that discipline. Cyber resilience is no exception.

**Disclaimer**

The views and conclusions contained herein are those of the authors and should not be interpreted as necessarily representing the official policies or endorsements, either expressed or implied, of their employers.

**References**

[1] A. Kott and I. Linkov, *Cyber Resilience of Systems and Networks*. Springer, 2019.

[2] A. Kott and P. Theron, "Doers, Not Watchers: Intelligent Autonomous Agents Are a Path to Cyber Resilience," *IEEE Secur. Priv.*, vol. 18, no. 3, pp. 62–66, 2020.

[3] M.-V. Florin and I. Linkov, *IRGC Resource Guide on Resilience*. Lausanne: EPFL International Risk Governance Center, 2016.

[4] I. Linkov, D. A. Eisenberg, K. Plourde, T. P. Seager, J. Allen, and A. Kott, "Resilience metrics for cyber systems," *Environ. Syst. Decis.*, vol. 33, no. 4, pp. 471–476, 2013.

[5] National Research Council, *Disaster resilience: A national imperative*. Washington, D.C.: The National Academies Press, 2012.

[6] A. Kott, J. Ludwig, and M. Lange, "Assessing Mission Impact of Cyberattacks: Toward a Model-Driven Paradigm," *IEEE Secur. Priv.*, vol. 15, no. 5, pp. 65–74, 2017.

[7] A. A. Ganin *et al.*, "Multicriteria Decision Framework for Cybersecurity Risk Assessment and Management," *Risk Anal.*, vol. 40, no. 1, pp. 183–199, 2020.

[8] D. S. Alberts and R. E. Hayes, *Understanding Command and Control*, vol. 61, no. 3. 2006.

[9] L. A. (Tony) Cox Jr., "What's wrong with risk matrices?," *Risk Anal.*, vol. 28, no. 2, pp. 497–512, 2008.

[10] D. W. Hubbard and R. Seiersen, *How to measure anything in cybersecurity risk*. Wiley, 2016.

[11] E. J. M. Colbert, A. Kott, and L. P. Knachel, "The game-theoretic model and experimental investigation of cyber wargaming," *J. Def. Model. Simul.*, vol. 17, no. 1, pp. 21–38, 2020.

[12] H. Abbass, A. Bender, S. Gaidow, and P. Whitbread, "Computational Red Teaming: Past, Present and Future," *IEEE Comput. Intell. Mag.*, no. February, pp. 30–42, 2011.

[13] L. Russo, F. Binaschi, and A. De Angelis, "Cybersecurity Exercises: Wargaming and Red Teaming," in *Next Generation CERTs*, A. Armando, M. Henauer, and A. Rigoni, Eds. IOS Press, 2019, pp. 44–60.

[14] S. Mansfield-Devine, "The best form of defence – the benefits of red teaming," *Comput. Fraud Secur.*, no. 10, pp. 8–12, 2018.

[15] S. Randhawa, B. Turnbull, J. Yuen, and J. Dean, "Mission-centric Automated Cyber Red Teaming," *ARES - 13th Int. Conf. Availability, Reliab. Secur.*, pp. 1–11, 2018.

[16] A. A. Ganin *et al.*, "Operational resilience: Concepts, design and analysis," *Sci. Rep.*, vol. 6, pp. 1–12, 2016.

[17]   I. Linkov and B. D. Trump, *The Science and Practice of Resilience*. Springer, 2019.

[18]   J. O. Kephart, G. B. Sorkin, M. Swimmer, and S. R. White, "Blueprint for a Computer Immune System," in *Virus Bulletin International Conference*, 1997.